\documentclass[usenatbib,longauth,traditabstract]{aa}

\usepackage{txfonts}
\usepackage{graphicx}
\usepackage{natbib}
 
\begin{document}
 
\title{CoRoT LRa02\_E2\_0121: Neptune-size planet candidate turns into a hierarchical triple system with a giant primary\thanks{Based on observations made with the 1.93-m telescope at Observatoire de Haute-Provence (CNRS), France, the 3.6-m telescope at La Silla Observatory (ESO), Chile (program 184.C-0639), the VLT at Paranal Observatory (ESO), Chile (program 083.C-0690), and the 2.1-m Otto Struve telescope at McDonald Observatory, Texas, USA.}}

\author{
L. Tal-Or \inst{1} 
\and A. Santerne \inst{2} 
\and T. Mazeh \inst{1}
\and F. Bouchy \inst{3,17} 
\and C.\ Moutou\inst{2} 
\and R.\ Alonso\inst{4} 
\and D.\ Gandolfi\inst{5} 
\and
 S.\ Aigrain \inst{6} \and 
 M.\ Auvergne\inst{7} \and 
 P.\ Barge\inst{2} \and 
 A.\ S.\ Bonomo\inst{2} \and 
 P.\ Bord{\'e}\inst{8} \and 
 H.\ Deeg\inst{9,18} \and 
 S.\ Ferraz-Mello\inst{10} \and 
 M.\ Deleuil\inst{2} \and 
 R.\ Dvorak\inst{11} \and 
 A.\ Erikson\inst{12} \and 
 M.\ Fridlund\inst{5} \and 
 M.\ Gillon\inst{6,19} \and 
 E.\ W.\ Guenther\inst{13} \and 
 T.\ Guillot\inst{14} \and 
 A.\ Hatzes\inst{13} \and 
 L.\ Jorda\inst{2} \and 
 H.\ Lammer\inst{15} \and 
 A.\ L{\'e}ger\inst{8} \and 
 A.\ Llebaria\inst{2} \and 
 M.\ Ollivier\inst{8} \and 
 M.\ P{\"a}tzold\inst{16} \and 
 D.\ Queloz\inst{4} \and 
 H.\ Rauer\inst{12,20} \and 
 D.\ Rouan\inst{7} \and 
 Y.\ Tsodikovich\inst{1} \and 
 G.\ Wuchterl\inst{13} 
}
 
\institute{School of Physics and Astronomy, Raymond and Beverly Sackler Faculty of
Exact Sciences, Tel Aviv University, Tel Aviv, Israel\\
\email{levtalo@post.tau.ac.il}
\and
Laboratoire d'Astrophysique de Marseille, Universit\'{e} de Provence \&
CNRS, 38 rue Fr\'{e}d\'{e}ric Joliot-Curie, 13388 Marseille cedex 13, France
\and
Observatoire de Haute Provence, CNRS/OAMP, 04870 St Michel l'Observatoire, France
\and
Observatoire de l'Universit\'{e} de Gen\`{e}ve, 51 chemin des Maillettes, 1290 Sauverny, Switzerland 
\and
Research and Scientific Support Department, ESTEC/ESA, Keplerlaan 1, 2200AG, Noordwijk, The Netherlands 
\and
Oxford Astrophyiscs, Denys Wilkinson Building, Keble Road, Oxford OX1 3RH, UK \and
LESIA, Observatoire de Paris, Place Jules Janssen, 92195 Meudon Cedex, France \and
Institut d'Astrophysique Spatiale, Universit\'{e} Paris XI, 91405 Orsay, France \and
Instituto de Astrof\'{\i}sica de Canarias, 38205 La Laguna, Tenerife, Spain \and
IAG, University of S\~{a}o Paulo, Brasil \and
University of Vienna, Institute of Astronomy, T\"{u}rkenschanzstr. 17, 1180 Vienna, Austria \and
Institute of Planetary Research, German Aerospace Center, Rutherfordstrasse 2, 12489 Berlin, Germany \and
Th\"{u}ringer Landessternwarte, Sternwarte 5, Tautenburg, 07778 Tautenburg, Germany \and
Universit\'{e} de Nice-Sophia Antipolis, CNRS UMR 6202, Observatoire de la C\^{o}te d'Azur, BP 4229, 06304 Nice Cedex 4, France \and
Space Research Institute, Austrian Academy of Science, Schmiedlstr. 6, 8042 Graz, Austria \and
Rheinisches Institut f{\"u}r Umweltforschung an der Universit{\"a}t zu K{\"o}ln, Aachener Strasse 209, 50931 K{\"o}ln, Germany \and
Institut d'Astrophysique de Paris, UMR7095 CNRS, Universit\'{e} Pierre
\& Marie Curie, 98bis Bd Arago, 75014 Paris, France \and
Universidad de La Laguna, Dept. de Astrof\'\i sica,  38200 La Laguna,
Tenerife, Spain \and
University of Li\`{e}ge, All\'{e}e du 6 ao\^{u}t 17, Sart Tilman, Li\`{e}ge 1, Belgium \and
Center for Astronomy and Astrophysics, TU Berlin, Hardenbergstr. 36, 10623 Berlin, Germany
}
 
\date{Received ... ; accepted ...}
\abstract{ 
This paper presents the case of CoRoT LRa02\_E2\_0121, which was initially classified as a Neptune-size transiting-planet candidate on a relatively wide orbit of $36.3$ days. Follow-up observations were performed with UVES, Sandiford, SOPHIE and HARPS. These observations revealed a faint companion in the spectra. To find the true nature of the system we derived the radial velocities of the faint companion using TODMOR --- a two-dimensional correlation technique, applied to the SOPHIE spectra. Modeling the lightcurve with EBAS we discovered a secondary eclipse with a depth of $\sim0.07\%$, indicating a diluted eclipsing binary. Combined MCMC modeling of the lightcurve and the radial velocities suggested that CoRoT LRa02\_E2\_0121 is a hierarchical triple system with an evolved G-type primary and an A-type:F-type grazing eclipsing binary. Such triple systems are difficult to discover.
}
\authorrunning{Tal-Or et al.}
\titlerunning{CoRoT LRa02\_E2\_0121: Hierarchical triple system with a giant primary}
\maketitle
 
\section{Introduction}
 
Since its launch, the CoRoT space mission \citep{baglin06, Aigrain2008, auvergne09} has obtained more than $10^5$ lightcurves of stars in more than 15 different fields that are spatially projected toward the center and anticenter of the Galaxy, in search of transiting exoplanets \citep{CoRoT2011EPJWC}. In order to confirm a new transiting planet, a candidate has to pass a sequence of tests \citep[e.g., ][]{Brown2003}. The tests include careful analysis of the lightcurve \citep{carpano09}, photometric follow-up observations \citep{deeg09}, and finally spectroscopic follow-up observations \citep{bouchy09}.
 
The main source of false positive alarms in the CoRoT sample are eclipsing binary systems in various configurations \citep{almenara09}. One possible strong indication of the binary nature of a system is the detection of a faint component in its observed spectrum. However, sometimes the stellar secondary is too faint to be noticed at first glance, and special tools are needed to detect it. Such a tool is TODMOR \citep[e.g.,][]{zm94, TODMOR, TODMOR2}, a two-dimensional correlation algorithm that was proved several times in the past to be an efficient tool for measuring the radial velocities ($=$\thinspace RVs) of the two components of a binary system, even with a relatively faint secondary \citep[e.g.,][]{mazeh95, mazeh97, torres95}.
 
\begin{table}
\caption{Coordinates and magnitudes of C0121.}
\begin{tabular}{lll}
\hline
\hline
Main identifiers & & \\
\hline
CoRoT ID & $110680825$ & \\
2MASS ID & $06520760-0526137$ & \\
\hline
Coordinates & & \\
\hline
RA (J2000) & $06^{h} 52^{m} 07^{s}.61$ & \\
Dec (J2000) & $-05^{\circ} 26' 13''.8 $ & \\
\hline
Filter & Magnitude & Source \\
\hline
\textit{B} & $12.32 \pm 0.05$ & \textit{ExoDat}$^a$ \\
\textit{V} & $11.75 \pm 0.08$ & \textit{ExoDat} \\
\textit{r'} & $11.37 \pm 0.06$ & \textit{ExoDat} \\
\textit{i'} & $10.95 \pm 0.07$ & \textit{ExoDat} \\
\textit{J} & $9.723 \pm 0.025$ & 2MASS$^b$ \\
\textit{H} & $9.197 \pm 0.020$ & 2MASS \\
\textit{K} & $9.102 \pm 0.021$ & 2MASS \\
\hline
\hline
$^a$ \citet{ExoDatDeleuil2009} & & \\
$^b$ \citet{2mass} & & \\
\end{tabular}
\label{MAG}
\end{table}
 
This paper presents the case of the transiting Neptune-size candidate LRa02\_E2\_0121 ($=$\thinspace C0121). By careful analysis of the CoRoT lightcurve combined with spectroscopic follow-up observations of this star, we concluded that it is a hierarchical triple system with an evolved G-type primary and a grazing eclipsing binary. The case is presented here to illustrate the long curvy road one might need to travel until the true nature of a system is revealed and to demonstrate the potential of TODMOR, which can reveal the diluted-binary nature of a system with a relatively modest investment of observational resources.
 
Section 2 gives some details about the star and presents the CoRoT white lightcurve. Section 3 describes the spectroscopic follow-up observations and the spectral analysis. Section 4 describes our application of TODMOR to the observed SOPHIE spectra. Section 5 presents our analysis of the CoRoT lightcurve and the combined MCMC modeling of the photometric and RV data. The astrophysics of C0121 as a hierarchical triple system with a G-type giant is detailed in Section 6. Section 7 presents some general conclusions derived from this specific case.
 
\section{The CoRoT lightcurve}
 
\begin{figure}
\resizebox{\hsize}{!}
{\includegraphics{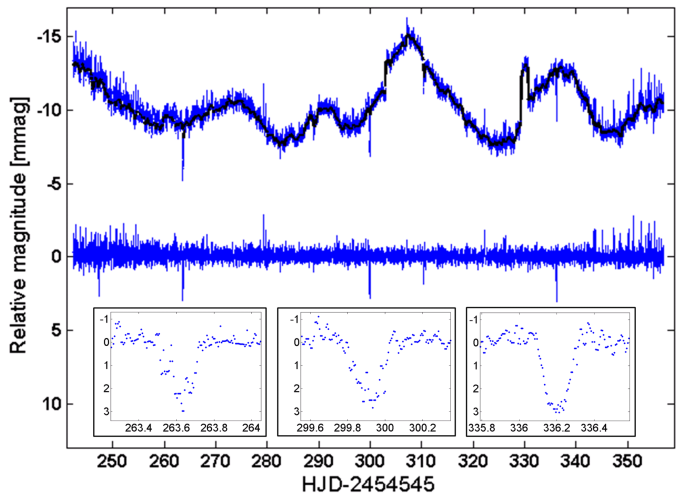}}
\caption{White lightcurve of C0121. Top: the calibrated lightcurve after rebinning oversampled points back to $512$\thinspace s and removing outliers. The calculated long-term trend is overplotted with a solid black line. Middle: the detrended lightcurve (for convenience the calibrated lightcurve was moved up by $10$\thinspace mmag). Insets: zoom on the three individual transit-like events.}
\label{LCO}
\end{figure}
 
The star C0121 was discovered as a transiting candidate by the alarm mode of CoRoT \citep{CoRoTalarm} during the LRa02 run, which lasted from Nov. 16, 2008 to Mar. 11, 2009. Its white lightcurve displayed three transit-like events of about $0.3\%$ depth with a period of about $36.3$ days, and therefore C0121 was identified as an interesting candidate of a Neptune-size planet with a relatively long period.
 
Figure \ref{LCO} shows the calibrated and detrended white lightcurves of this candidate, with a zoom on the three transit-like events in the insets. Trends were removed with local regression that fits a second-degree polynomial to one-day-long data with weighted linear least squares. Table \ref{MAG} lists some basic information on C0121. 
 
\section{Spectroscopic follow-up observations}

\subsection{Radial velocity measurements}
 
The first five RV measurements of C0121 were obtained with the SOPHIE spectrograph \citep{bouchy06, SOPHIE}, mounted at the $1.93$\thinspace m telescope at the Observatoire de Haute Provence, France, between Feb. 26 and Apr. 04, 2009. Following the discovery of the transits in the alarm mode, the observations began before the end of the LRa02 run and spanned $37$ days to cover a full orbital period of the transiting object. The observation timings were chosen to be close to the quadrature times of the transiting object to maximize the chance that small RV variability will be detected. These five observations gave us the first signs of the complexity of the system. Instead of the expected $36.3$-day periodicity, we detected a long-term drift, with an amplitude of about $0.5$ km\thinspace s$^{-1}$.
 
\begin{figure}
\resizebox{\hsize}{!}
{\includegraphics{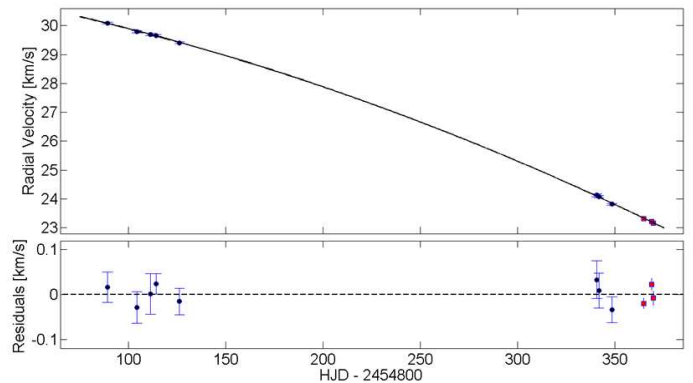}}
\caption{RVs of the primary in C0121 as measured by the SOPHIE (black circles) and HARPS (red boxes) spectrographs using CCF. The solid black line is a second-degree polynomial fit. The residuals are plotted at the bottom.}
\label{PPP}
\end{figure}

Six more observations of C0121 were obtained when the field was observable again, between Nov. 05 and Dec. 04, 2009. Three were acquired with SOPHIE and three other ones with the HARPS spectrograph \citep{pepe00, mayor03}, mounted at the ESO $3.6$\thinspace m telescope in La Silla Observatory, Chile. All spectra had a typical signal-to-noise ratio (S/N) of $30$--$50$ per pixel at $550$\thinspace nm. In the additional six spectra we found the same long-term drift, which at that time had an observed amplitude of about $7$ km\thinspace s$^{-1}$. The long-term variation indicated that C0121 is a member of a binary system with a long period, but did not rule out the existence of a planet in the system. There could still be a planet orbiting one of the members of the wide binary, as in the cases of 16 Cyg B \citep{16CygB}, $\gamma$ Cephei A \citep{Hatzes2003}, or 30 Ari B \citep{30AriB}. Obviously, one needs more observations to understand such a complex system. Figure \ref{PPP} shows the RVs of the primary, indicating that the orbital period is too long for us to derive an orbit, so we fit the measured RVs with a second-degree polynomial.
 
The plot started to thicken when we noticed that the three spectra taken with HARPS revealed a faint secondary. Following this, we re-analyzed the eight spectra already obtained with SOPHIE, fitting a sum of two Gaussians to the cross correlation function \citep[$=$\thinspace CCF,][]{baranne96} derived with a G2 mask \citep{Pepe2002}. Due to the low S/N of the secondary feature in the CCF, we succeeded in measuring its RV in only 9 out of the 11 measured spectra --- three from HARPS and six from SOPHIE spectra. All measured RVs are listed in Table \ref{RVT}.
 
The nine measured velocities of the faint secondary were consistent with the $36.3$-day period found by CoRoT, but with a large RV semiamplitude of $\sim62$ km\thinspace s$^{-1}$. This finding ruled out completely the existence of a transiting planet in C0121, and indicated that the extremely shallow transit-like signal is coming from its faint companion, which is by itself a diluted eclipsing binary.
 
Along the way we found one more surprise: the measured RVs of the faint companion of C0121 were in \textit{anti-phase} with the CoRoT lightcurve; i.e., the detected eclipses occurred on the rising part of the RV curve. This means that the eclipses detected in the CoRoT lightcurve are \textit{caused} by the star whose spectrum is detected in the composite spectra of the system, passing in front of its companion. The spectrum of the {\it eclipsed} star is not seen in the observed spectra. This is a very unusual case, because in most eclipsing binaries the primary eclipse occurs when the primary star, whose spectrum dominates the spectra, is eclipsed. 

It was clear then that C0121 includes at least three stars. To facilitate the discussion we denoted the primary star in the spectra by C0121-A, the secondary star in the spectra, whose RV was observed to vary with a large amplitude by C0121-Bb, and the {\it eclipsed} star during the primary eclipses detected so far by C0121-Ba.


\begin{table}
\caption{The measured RVs of C0121 (km\thinspace s$^{-1}$).}
\begin{tabular}{cccc}
\hline
\hline
HJD & CCF: & CCF: & S/N Per Pixel \\
($-2450000$) & Primary & Secondary & (at $550$\thinspace nm) \\
\hline
$ 4889.30773 $ & $ 30.091 \pm 0.034 $ & $  70.07 \pm 1.78 $ & $33.9$ \\
$ 4904.29007 $ & $ 29.789 \pm 0.035 $ & $ --- $ & $35.1$ \\
$ 4911.38470 $ & $ 29.694 \pm 0.045 $ & $ -45.34 \pm 1.76 $ & $27.8$ \\
$ 4914.30337 $ & $ 29.663 \pm 0.023 $ & $ -21.05 \pm 1.34 $ & $49.8$ \\
$ 4926.31911 $ & $ 29.403 \pm 0.030 $ & $  74.39 \pm 2.05 $ & $43.3$ \\
$ 5140.63895 $ & $ 24.128 \pm 0.042 $ & $ --- $ & $29.4$ \\
$ 5141.70360 $ & $ 24.071 \pm 0.039 $ & $  72.67 \pm 2.52 $ & $31.3$ \\
$ 5148.56025 $ & $ 23.815 \pm 0.029 $ & $  62.39 \pm 2.15 $ & $37.8$ \\
$ 5164.81355 $ & $ 23.312 \pm 0.012 $ & $ -41.98 \pm 0.87 $ & $51.0$ \\
$ 5168.84324 $ & $ 23.225 \pm 0.014 $ & $ -11.56 \pm 0.74 $ & $48.4$ \\
$ 5169.70123 $ & $ 23.166 \pm 0.016 $ & $  -0.86 \pm 0.73 $ & $43.8$ \\
\hline
HJD & TODMOR: & TODMOR:  & S/N Per Pixel \\
($-2450000$) & Primary & Secondary  & (at $550$\thinspace nm)\\
\hline
$ 4889.30773 $ & $ 30.11 \pm 0.11 $ & $  74.08 \pm 1.32 $  & $33.9$ \\
$ 4904.29007 $ & $ 29.82 \pm 0.12 $ & $ -27.05 \pm 1.14 $  & $35.1$ \\
$ 4911.38470 $ & $ 29.72 \pm 0.19 $ & $ -44.60 \pm 1.77 $  & $27.8$ \\
$ 4914.30337 $ & $ 29.68 \pm 0.10 $ & $ -22.13 \pm 0.89 $  & $49.8$ \\
$ 4926.31911 $ & $ 29.40 \pm 0.12 $ & $  75.61 \pm 1.20 $  & $43.3$ \\
$ 5140.63895 $ & $ 23.93 \pm 0.11 $ & $  68.04 \pm 1.33 $  & $29.4$ \\
$ 5141.70360 $ & $ 23.85 \pm 0.11 $ & $  73.55 \pm 1.28 $  & $31.3$ \\
$ 5148.56025 $ & $ 23.89 \pm 0.10 $ & $  65.67 \pm 1.20 $  & $37.8$ \\
\hline
\hline
\end{tabular}
\label{RVT}
\end{table}
 
\subsection{High-resolution spectroscopic observations}
 
A reconnaissance high-resolution ($R\approx47\,000$) spectrum of C0121 was acquired on February 13, 2009, using the Sandiford Cassegrain echelle spectrograph \citep{SANDIFORD} mounted at the $2.1$\thinspace m Otto Struve Telescope at McDonald Observatory. The analysis of the spectrum revealed a G-type giant star with $T_{\rm eff}=5500\pm250$\thinspace K, log\thinspace $g=3.0\pm0.2$, and $v\sin i=12\pm1$ km\thinspace s$^{-1}$.
 
Furthermore, high-resolution, high-S/N spectroscopy of C0121 was performed with the UVES spectrograph \citep{UVES} mounted at the ESO $8.2$\thinspace m Very Large Telescope ($=$\thinspace VLT) in Paranal Observatory, Chile. Two consecutive exposures of $3000$\thinspace sec each were acquired on April 8, 2009, under the ESO program 083.C-0690. The adopted set-up yielded a resolving power of $R\approx67\,000$ in the spectral range of $3280-6820$~\AA, with a final S/N of about $320$ per pixel at $5500$~\AA. Following the procedure already adopted in some CoRoT discovery papers \citep[e.g.,][]{CoRoT3b,CoRoT11b}, the UVES spectrum was used to derive the fundamental photospheric parameters of C0121. Ignoring the additional light from C0121-Ba and C0121-Bb, we found $T_{\rm eff}=5640\pm100$\thinspace K, log\thinspace $g=2.9\pm0.1$, $[\rm{M/H}]=-1.0\pm0.2$, and $v\sin i=11.5\pm1.0$ km\thinspace s$^{-1}$, corresponding to a G0\,III star. Allowing a dilution factor and fitting only the Mg\,I, Fe\,I/II and Ca\,I lines we found $T_{\rm eff}=4980\pm100$\thinspace K, log\thinspace $g=2.6\pm0.1$, $[\rm{M/H}]=-0.5\pm0.2$, and $v\sin i=11.0\pm1.0$ km\thinspace s$^{-1}$, corresponding to a G8\,III star.
 
The two observations indicate that C0121-A is probably a G-type giant star.

\section{TODMOR analysis of the SOPHIE spectra}
 
To better understand the unusual C0121 system, we analyzed all eight SOPHIE spectra with our two-dimensional correlation algorithm TODMOR, and derived the RVs of C0121-Bb from all eight spectra. Our best model included as the primary template the observed spectrum of HD\,32923, a G4V star \citep{HD32923}, rotationally broadened by $9$\thinspace km\thinspace s$^{-1}$, and as the secondary template the observed spectrum of HD\,185395 ($\theta$ Cyg), an F4V star \citep{HD185395}. The  secondary/primary flux ratio was $\sim0.07$.
 
Despite our efforts, we did not detect any spectral signature of C0121-Ba, even when analyzing the data with TRIMOR \citep{TRIMOR}, an algorithm to analyze multi-order spectra of triple systems. This might indicate that Ba is a substantially hotter star, with spectral type A or even earlier. Such stars have only few broad spectral lines in the optical band, and therefore it is hard to measure their RVs, especially when their light is blended with the light of other stars.
 
\begin{figure}
\resizebox{\hsize}{!}
{\includegraphics{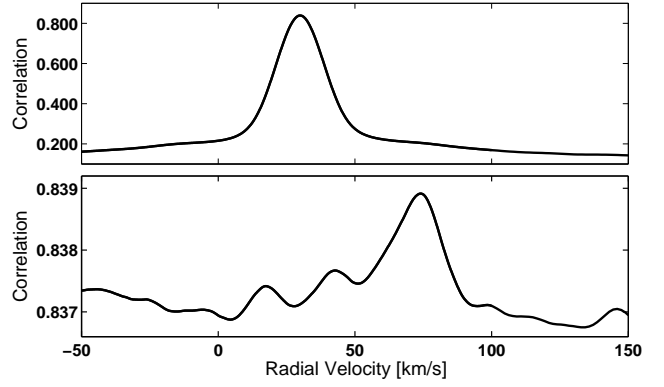}}
\caption{TODMOR plot for the first exposure of C0121. The upper and lower panels show the primary and secondary
\textit{cuts} through the two-dimensional correlation function.}
\label{TTT}
\end{figure}
 
Figure \ref{TTT} shows the TODMOR results for the first SOPHIE exposure of C0121. The upper and lower panels of the figure show primary and secondary \textit{cuts} through the two-dimensional correlation function that run through the correlation peak. The primary cut (upper panel) is parallel to the primary RV axis, freezing the secondary velocity at its derived velocity.  The secondary cut (lower panel) runs parallel to the secondary RV axis, freezing the primary velocity at its derived velocity.
 
We note that in the upper panel the correlation drops by $\sim0.6$ when moving away from the peak, because we change the velocity of the \textit{primary} template in the model. On the other hand, the correlation in the lower panel drops only by $\sim0.002$, because we change the velocity of the \textit{secondary} template, which contributes only $\sim7$\% of the light, and the primary velocity is kept at its best value. Nevertheless, the peak at the lower panel is prominent, indicating the high significance of the detection of the secondary velocity.
 
\begin{figure}
\resizebox{\hsize}{!}
{\includegraphics{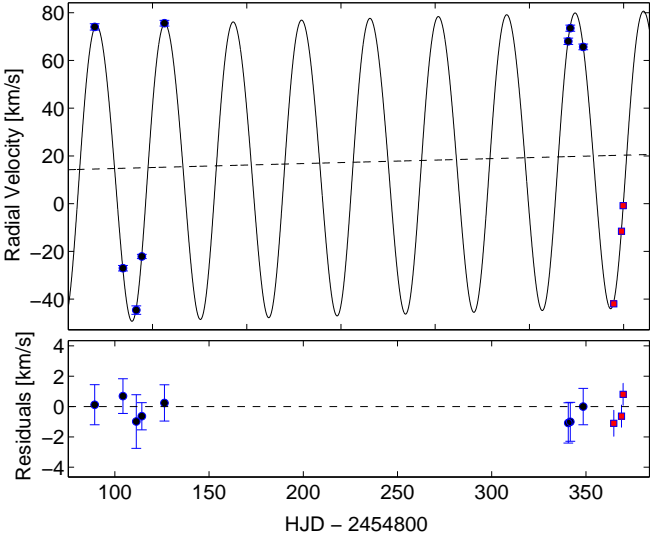}}
\caption{RVs of C0121-Bb as measured by the SOPHIE (black circles) and HARPS (red boxes) spectrographs. SOPHIE RVs were derived with TODMOR, while HARPS RVs were derived by fitting a sum of two Gaussians to the CCF. The solid line is the Keplerian model produced from the orbital parameters in Table \ref{PAR} and the dashed line is the center-of-mass velocity. The residuals are plotted at the bottom.}
\label{SSS}
\end{figure}
 
The TODMOR RVs of C0121-A and C0121-Bb are listed in Table \ref{RVT}. Figure \ref{SSS} presents the RVs of C0121-Bb with our best orbital solution, which is described in the next section.
 
We also estimated the effective temperatures of C0121-A and C0121-Bb from the SOPHIE spectra by trying a range of high-resolution PHOENIX synthetic models \citep{phoenix99} with different $T_{\rm eff}$ values as templates. TODMOR found $T_{\rm eff,A}=5650\pm200$\thinspace K and $T_{\rm eff,Bb}=6500\pm500$\thinspace K, in agreement with the analysis of the UVES and Sandiford observations.
 
\section{Modeling the eclipsing binary C0121-B}

\begin{figure}
\resizebox{\hsize}{!}
{\includegraphics{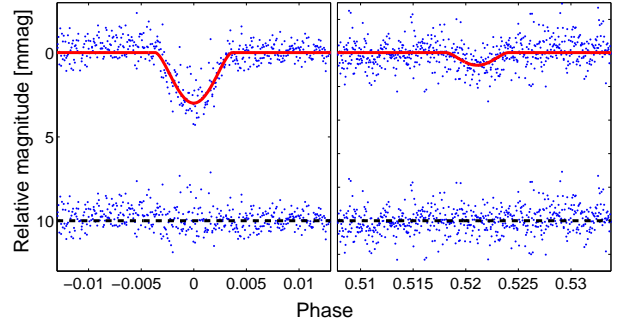}}
\caption{Zoom on the primary and secondary eclipses in the phase-folded lightcurve of C0121 with our best model overplotted with a solid red line. The residuals were moved down by $10$\thinspace mmag for convenience.}
\label{LCM}
\end{figure}

To estimate the photometric parameters of C0121-B we re-analyzed the CoRoT white lightcurve with EBAS \citep{EBAS,EBAS2}, an algorithm for analyzing eclipsing-binary lightcurve based on the EBOP subroutines \citep{Popper1981}. Our EBAS analysis discovered a faint secondary eclipse with a depth of $(0.07\pm 0.01)\%$ at the orbital phase $\sim0.52$, showing again that C0121-B is actually an eclipsing binary (see Figure \ref{LCM}). The CoRoT lightcurve covered three primary and four secondary eclipses.
 
\begin{table*}
\caption{The parameters of C0121-B from the combined lightcurve and RV modeling.}
\centering
\begin{tabular}{llrl}
\hline
\hline
Symbol & Parameter name & Value & Units\\
\hline
& Orbital parameters & & \\
\hline
$ P $ & Period & $ 36.2873 \pm 0.0025 $ & day \\
$ e\cos\omega $ & Eccentricity $\times$ Cosine Longitude of periastron & $ 0.0331 \pm 0.0003 $ & -- \\
$ e\sin\omega $ & Eccentricity $\times$ Sine Longitude of periastron & $ 0.029 _{-0.007} ^{+0.008} $ & -- \\
$ K $ & RV semiamplitude & $ 62.2 \pm 0.5 $ & km\thinspace s$^{-1}$ \\
$ \gamma $ & Center-of-mass RV at $JD=2454889$ & $ 14.5 \pm 0.6 $ & km\thinspace s$^{-1}$ \\
$ \dot{\gamma} $ & Center-of-mass acceleration & $ 20.6 \pm 2.8 $ & m\thinspace s$^{-1}$d$^{-1}$ \\
$ T $ & Time of periastron (calculated analytically) & $ 2454876.7 \pm 0.7 $ & JD \\
\hline
& Photometric parameters & & \\
\hline
$ T_0 $ & Time of center of primary eclipse & $ 2454844.9152 \pm 0.0020 $ & JD \\
$ J_s $ & Surface-brightness ratio (secondary/primary) & $ 0.61 \pm 0.23 $ & -- \\
$ r_t $ & Fractional sum of radii & $ 0.049 \pm 0.006 $ & -- \\
$ k $ & Ratio of radii (secondary/primary) & $ 0.60 \pm 0.25 $ & -- \\
$ x $ & Impact parameter & $ 0.88 _{-0.04} ^{+0.02} $ & -- \\
$ L_3 $ & Third-light (blending) fraction & $ 0.86 \pm 0.04 $ & -- \\
$ u_p $ & Limb-darkening coefficient of primary (fixed) & $ 0.55 $ & -- \\
$ u_s $ & Limb-darkening coefficient of secondary (fixed) & $ 0.58 $ & -- \\
\hline
& Other estimated parameters  & & \\
\hline
$ i $ & inclination (see Eq.(2)) & $ 87.45 _{-0.35} ^{+0.43} $ & degree \\
$ a_{Bb} $ & Semimajor axis of secondary & $ 0.2076 \pm 0.0017 $ & AU \\
$ f $ & Mass function of Bb& $ 0.903 \pm 0.022 $ & M$_{\odot}$ \\
$ q $ & Mass ratio (M$_{Ba}$/M$_{Bb}$) assuming M$_{Bb}=1.3$M$_{\odot}$ & $ 1.73 \pm 0.03 $ & -- \\
$F_{Bb}$:$F_{Ba} $ & The Bb:Ba flux ratio & $0.6_{-0.5}^{+0.3}$ & -- \\
\hline
\hline
\end{tabular}
\label{PAR}
\end{table*}
 
Final parameters were derived with MCMC analysis \citep[e.g.,][]{Tegmark,Ford2005AJ} of the white lightcurve, {\it together} with the RV data. The input data were the detrended white lightcurve from CoRoT, which included $19,348$ points and the $11$ measured RVs of C0121-Bb. From the RVs listed in Table \ref{RVT} we used the $8$ TODMOR-derived SOPHIE RVs and the $3$ CCF-derived HARPS RVs. Errors for the photometric data were derived from the local scatter of the detrended lightcurve (see Figure \ref{LCO}).
 
The sizes of the MCMC Gaussian perturbations were set by the error estimations of EBAS and by a relatively short MCMC run (of $\sim10^5$ accepted steps). Following the Metropolis-Hastings algorithm, trial points with lower $\chi^{2}$ were accepted, whereas trial points with higher $\chi^{2}$ were accepted only with probability of exp($-\Delta\chi^{2}/2$), assuming the observational errors to be Gaussian \citep{Ford2005AJ}. The $\chi^{2}$ for each trial point was the sum of the lightcurve $\chi^{2}$ and the $\chi^{2}$ of the RV data. We used flat prior distributions for all free parameters of the model, all detailed in Table~3. The absolute values of $e\cos\omega$, $e\sin\omega$, $J_s$, $r_t$, $k$, $x$ and $L_3$ were constrained to be between $0$ and $1$.
 
The fractional sum of radii ($r_t$) and the impact parameter ($x$) can be expressed by the more conventional parameters as
\begin{equation}
r_t=\frac{R_p+R_s}{a} 
\end{equation}
and 
\begin{equation}
x=\frac{\cos i}{r_t}\frac{1-e^2}{1+e\sin\omega} $ , $ 
\end{equation}
where $R_p$ and $R_s$ are the primary and secondary radii, $a$ is the orbital semimajor axis, $i$ the inclination, $e$ the eccentricity, and $\omega$ the longitude of periastron \citep{EBAS}. Using this convention,  $x=1$ when the components are grazing but not yet eclipsing. We used linear limb darkening law, with coefficients, $u_p$ and $u_s$, fixed according to \citet{Sing2010}, assuming an A5 and F5 dwarfs for the primary and secondary stars in the eclipsing binary.
 
The MCMC run consisted of $2\,10^6$ accepted steps. The derived value of each parameter was its MCMC median, and its confidence limits were estimated as the range of MCMC values that cover the central $68.3$\% of the chain.
 
Table \ref{PAR} gives the full list of the estimated parameters of C0121-B from the MCMC analysis. Most of the model parameters showed normal posterior probability distribution. For the ones with skewed distribution, Table \ref{PAR} gives uneven confidence limits. 
Figure \ref{LCM} presents the folded lightcurve and our best model,  zoomed on the primary and secondary eclipses.
 
We note that the high $x$ value we found indicates that the C0121-B eclipsing binary is grazing. In addition, the significantly non-zero center-of-mass acceleration value we found indicates that the C0121-B binary is itself a member of a hierarchical triple system.
 
The time of periastron passage ($T$), as well as some other parameters of C0121-B, which are listed at the bottom of Table \ref{PAR}, were analytically calculated for each point in the MCMC chain. Their values and errors were estimated the same way as for the free parameters of the model. Only for the Bb:Ba flux ratio did we take the \textit{most probable} value of $J_s\,k^2$ as our estimation, where $J_s$ is the Bb:Ba surface-brightness ratio and $k$ their ratio of radii (see Table \ref{PAR}), and not the median one, because we found its probability distribution to be extremely skewed.
 
C0121 was bright enough for CoRoT to observe it with its three different bands \citep[e.g.,][]{Aigrain2008, CoRoT8b}. The reduction of the colored lightcurves, which was done after the completion of the LRa02 run, revealed different depths of the primary eclipse --- $(0.42\pm 0.02)\%$ in blue, $(0.35\pm 0.02)\%$ in green, and $(0.18\pm 0.02)\%$ in red, indicating again that C0121-Ba is probably hotter than the other two stars in the system.
 
\section{The C0121 system}
 
\begin{figure}
\resizebox{\hsize}{!}
{\includegraphics{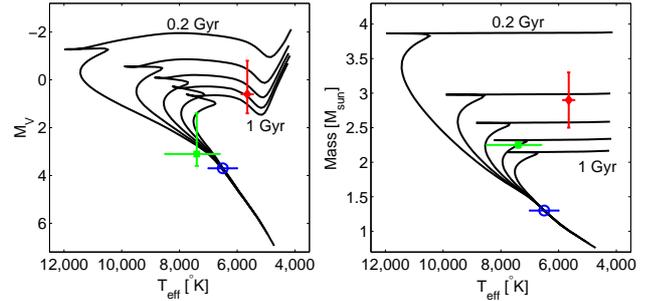}}
\caption{Y$^2$ stellar isochrones from \citet{Demarque2004} for solar metallicity ($[$Fe/H$]=0.05$, $[\alpha$/Fe$]=0$) and ages of $0.2,0.4,0.6,0.8$, and $1$ Gyr, from top-left to bottom-right. Left: M$_V($T$_{\rm eff})$ isochrones. Right: M$($T$_{\rm eff})$ isochrones. Our estimations for C0121-A, C0121-Ba, and C0121-Bb are marked by a red diamond, a green square, and a blue circle, respectively. The absolute magnitude and mass of C0121-Bb were taken as the average value of the isochrones for a $6500$\thinspace K star, assuming no error.}
\label{ISO}
\end{figure}

The light of C0121 is coming from (at least) three stellar sources: a G-type star (C0121-A) on a wide orbit with an unknown period and an eclipsing binary (C0121-B) that includes an F-type star (C0121-Bb) orbiting a hotter and more massive star (C0121-Ba) on a nearly circular orbit of $\sim36.3$ days. In addition, the center of mass of the C0121-B system is itself on a wide orbit with unknown period. We explore now the assumption that the three observed stars, A, Ba, and Bb, consist of a \textit{single} hierarchical triple system. To do that, we use the following previously-estimated quantities and relations, derived from the {\it white} lightcurve and spectra of C0121:
 
\begin{itemize}
 \item The effective temperature of C0121-Bb is $6500\pm500$;
 \item The effective temperature of C0121-A is $5650\pm200$;
 \item The Bb:Ba surface-brightness ratio is $0.61\pm0.23$;
 \item The Bb:Ba flux ratio is $0.6_{-0.5}^{+0.3}$;
 \item C0121-A contributes $86\pm4$\% of the total light of the system;
 \item The Ba:Bb mass ratio $q$ depends on $M_{Bb}$ through ($M_{Bb}f^{-1}\sin^3i$)\thinspace $q^3-q^2-2q-1=0$, where 
the mass function $f$ is $0.903\pm0.022$\thinspace M$_{\odot}$ and the inclination $i$ is $87.4\pm0.4^{\circ}$;
 \item The center-of-mass acceleration of C0121-B is $20.6\pm2.8$\thinspace m\thinspace s$^{-1}$d$^{-1}$;
 \item The average acceleration value of C0121-A is $-25\pm1$\thinspace m\thinspace s$^{-1}$d$^{-1}$.
\end{itemize}
To proceed, we {\it assume} that Bb is on the main sequence. From Y$^2$ stellar isochrones of solar-metallicity stars \citep{Demarque2004} we take for it the following values (see Fig. \ref{ISO}):
\begin{itemize}
 \item The $M_V$ brightness of Bb is $3.7$;
 \item The mass of Bb is $1.3$\thinspace M$_{\odot}$.
\end{itemize}
 
To estimate $T_{\rm eff,Ba}$ we used the derived surface brightness ratio and calculated the expected surface-brightness ratio between various PHOENIX synthetic models, taking the CoRoT total efficiency \citep{CoRoTResponse} into account. We got $T_{\rm eff,Ba}=7400_{-800}^{+1100}$\thinspace K.
From the assumed $M_V$ value of Bb, and using the derived brightness ratios, we  got for C0121-Ba and C0121-A an $M_V$ values of $3.1_{-1.7}^{+0.5}$ and $0.6_{-1.4}^{+0.8}$, respectively.
From the assumed mass of Bb and the derived ratios, including the acceleration ratio between A and B, we got for C0121-Ba and C0121-A masses of $2.25\pm0.04$\thinspace M$_{\odot}$ and $2.8\pm0.4$\thinspace M$_{\odot}$, respectively.
 
To see whether the three temperatures, brightnesses, and masses can be fitted on a single isochrone, we plotted the Y$^2$ stellar isochrones for solar metallicity and ages of $0.2$--$1$ Gyr in Figure \ref{ISO}, together with our estimations for the effective temperatures, masses, and $M_V$ values of the three observed stars. It can be seen that a single isochrone of $\sim0.6$ Gyr could fit all nine estimated values, supporting the hierarchical triple hypothesis.
 
Another way to estimate the stellar temperatures is to use the lightcurves of the different CoRoT bands. We derived the six $J_{s,j}$ and $L_{3,j}$ values, where $j$ indicates the color, using MCMC as before, and searched for the best temperatures of the three stars, together with the best Ba:A ratio of radii, $k_{Ba:A}$, similar to the way we estimated $T_{\rm eff,Ba}$ from the white lightcurve parameters. To do that we had to estimate the wavelength range of the blue, green, and red CoRoT bandpasses. From the observed magnitudes of C0121 in each channel, we estimated the two \textit{effective} wavelengths that separate the channels in the CoRoT mask to be roughly at $5100$ and $5600$~\AA. We got $T_{\rm eff,A}=5700\pm200$\thinspace K, $T_{\rm eff,Ba}=7200\pm400$\thinspace K, $T_{\rm eff,Bb}=6500\pm300$\thinspace K, and $k_{Ba:A}=0.20\pm0.02$. These values agree with the above estimated values and confirm that C0121-A is indeed a G-type giant star. 
 
The visual and absolute magnitudes of C0121-A yield a distance of $1.7_{-0.6}^{+1.7}$\thinspace kpc to the system, consistent with being in the Perseus arm of the Galaxy \citep{Russeil2003}.
 
\section{Conclusions}
 
The available photometric and spectral information for C0121 indicates that the system is a hierarchical triple system and that the distant companion, C0121-A, is a G-type giant star. Hierarchical triple systems with the distant component in its giant phase are quite difficult to discover. Even if the close pair is an eclipsing binary, the dilution factor makes the eclipse rather shallow. In our case the (grazing) eclipse depth is only $\sim 0.3$\%. Very recently, \citet{HD181068} discovered a similar case, HD\,181068, for which the observed depth of the close-pair eclipse was also much less than 1\%. Not surprisingly, the triple nature of HD\,181068 was also discovered by a lightcurve obtained from space, this time with the {\it Kepler} mission. Ground-based observations could not have detected the eclipses of the close binary in these systems. When the close pair is not eclipsing it is almost impossible to detect the triple nature of the system if the distant companion is a giant, even if the wide orbital period is short enough to be discovered by radial-velocity observations of the giant. The discoveries of these two systems indicate that triple systems might not be rare, as already found by \citet{mm87}, \citet{Tokovinin2008}, and \citet{Raghavan2010}.

The present case demonstrates how a promising Neptune-size transiting-planet candidate turned out to be a grazing eclipsing binary in a hierarchical triple system. Indeed, hierarchical triple systems are expected to be the main source for false positive detections of Neptune-size transiting-planet candidates in the \textit{Kepler} data too \citep{Morton2011}. The true nature of the system could have been revealed by analyzing the first few SOPHIE spectra with TODMOR, suggesting that an automated TODMOR pipeline could be useful in the long planet-hunting procedure. We hope to construct such a pipeline in the near future, so we can detect disguised hierarchical triple systems as early as possible.
 
\begin{acknowledgements}
We thank Gil Sokol, Gil Nachmani, and Simchon Faigler for their help with the lightcurve analysis. This research was supported by the Israel Science Foundation (grant No. 655/07). Financial support for the SOPHIE observations from the ``Programme National de Plan\'etologie'' (PNP) of CNRS/INSU, France is gratefully acknowledged. We also acknowledge support from the French National Research Agency (ANR-08-JCJC-0102-01). The German CoRoT Team acknowledges the support of DLR grants 50OW204 and 50OW603. We thank the anonymous referee for fruitful comments and suggestions.
\end{acknowledgements}
 
\bibliographystyle{aa}
\bibliography{c0121.bib}

\begin{thebibliography}{50}
\expandafter\ifx\csname natexlab\endcsname\relax\def\natexlab#1{#1}\fi

\bibitem[{{Aigrain} {et~al.}(2008){Aigrain}, {Barge}, {Deleuil}, {Fressin},
  {Moutou}, {Queloz}, {Auvergne}, \& {Baglin}}]{Aigrain2008}
{Aigrain}, S., {Barge}, P., {Deleuil}, M., {et~al.} 2008, in Astronomical
  Society of the Pacific Conference Series, Vol. 384, 14th Cambridge Workshop
  on Cool Stars, Stellar Systems, and the Sun, ed. {G.~van Belle}, 270--280

\bibitem[{{Almenara} {et~al.}(2009){Almenara}, {Deeg}, {Aigrain}, {Alonso},
  {Auvergne}, {Baglin}, {Barbieri}, {Barge}, {Bord{\'e}}, {Bouchy}, {Bruntt},
  {Cabrera}, {Carone}, {Carpano}, {Catala}, {Csizmadia}, {de La Reza},
  {Deleuil}, {Dvorak}, {Erikson}, {Fridlund}, {Gandolfi}, {Gillon}, {Gondoin},
  {Guenther}, {Guillot}, {Hatzes}, {H{\'e}brard}, {Jorda}, {Lammer},
  {L{\'e}ger}, {Llebaria}, {Loeillet}, {Magain}, {Mayor}, {Mazeh}, {Moutou},
  {Ollivier}, {P{\"a}tzold}, {Pont}, {Queloz}, {Rauer}, {R{\'e}gulo}, {Renner},
  {Rouan}, {Samuel}, {Schneider}, {Shporer}, {Wuchterl}, \&
  {Zucker}}]{almenara09}
{Almenara}, J.~M., {Deeg}, H.~J., {Aigrain}, S., {et~al.} 2009, \aap, 506, 337

\bibitem[{{Auvergne} {et~al.}(2009){Auvergne}, {Bodin}, {Boisnard}, {Buey},
  {Chaintreuil}, {Epstein}, {Jouret}, {Lam-Trong}, {Levacher}, {Magnan},
  {Perez}, {Plasson}, {Plesseria}, {Peter}, {Steller}, {Tiph{\`e}ne}, {Baglin},
  {Agogu{\'e}}, {Appourchaux}, {Barbet}, {Beaufort}, {Bellenger}, {Berlin},
  {Bernardi}, {Blouin}, {Boumier}, {Bonneau}, {Briet}, {Butler}, {Cautain},
  {Chiavassa}, {Costes}, {Cuvilho}, {Cunha-Parro}, {de Oliveira Fialho},
  {Decaudin}, {Defise}, {Djalal}, {Docclo}, {Drummond}, {Dupuis}, {Exil},
  {Faur{\'e}}, {Gaboriaud}, {Gamet}, {Gavalda}, {Grolleau}, {Gueguen},
  {Guivarc'h}, {Guterman}, {Hasiba}, {Huntzinger}, {Hustaix}, {Imbert},
  {Jeanville}, {Johlander}, {Jorda}, {Journoud}, {Karioty}, {Kerjean},
  {Lafond}, {Lapeyrere}, {Landiech}, {Larqu{\'e}}, {Laudet}, {Le Merrer},
  {Leporati}, {Leruyet}, {Levieuge}, {Llebaria}, {Martin}, {Mazy}, {Mesnager},
  {Michel}, {Moalic}, {Monjoin}, {Naudet}, {Neukirchner}, {Nguyen-Kim},
  {Ollivier}, {Orcesi}, {Ottacher}, {Oulali}, {Parisot}, {Perruchot},
  {Piacentino}, {Pinheiro da Silva}, {Platzer}, {Pontet}, {Pradines},
  {Quentin}, {Rohbeck}, {Rolland}, {Rollenhagen}, {Romagnan}, {Russ}, {Samadi},
  {Schmidt}, {Schwartz}, {Sebbag}, {Smit}, {Sunter}, {Tello}, {Toulouse},
  {Ulmer}, {Vandermarcq}, {Vergnault}, {Wallner}, {Waultier}, \&
  {Zanatta}}]{auvergne09}
{Auvergne}, M., {Bodin}, P., {Boisnard}, L., {et~al.} 2009, \aap, 506, 411

\bibitem[{{Baglin} {et~al.}(2006){Baglin}, {Auvergne}, {Boisnard}, {Lam-Trong},
  {Barge}, {Catala}, {Deleuil}, {Michel}, \& {Weiss}}]{baglin06}
{Baglin}, A., {Auvergne}, M., {Boisnard}, L., {et~al.} 2006, in 36th COSPAR
  Scientific Assembly, Vol.~36, 3749

\bibitem[{{Baranne} {et~al.}(1996){Baranne}, {Queloz}, {Mayor}, {Adrianzyk},
  {Knispel}, {Kohler}, {Lacroix}, {Meunier}, {Rimbaud}, \& {Vin}}]{baranne96}
{Baranne}, A., {Queloz}, D., {Mayor}, M., {et~al.} 1996, \aaps, 119, 373

\bibitem[{{Bord{\'e}} {et~al.}(2010){Bord{\'e}}, {Bouchy}, {Deleuil},
  {Cabrera}, {Jorda}, {Lovis}, {Csizmadia}, {Aigrain}, {Almenara}, {Alonso},
  {Auvergne}, {Baglin}, {Barge}, {Benz}, {Bonomo}, {Bruntt}, {Carone},
  {Carpano}, {Deeg}, {Dvorak}, {Erikson}, {Ferraz-Mello}, {Fridlund},
  {Gandolfi}, {Gazzano}, {Gillon}, {Guenther}, {Guillot}, {Guterman}, {Hatzes},
  {Havel}, {H{\'e}brard}, {Lammer}, {L{\'e}ger}, {Mayor}, {Mazeh}, {Moutou},
  {P{\"a}tzold}, {Pepe}, {Ollivier}, {Queloz}, {Rauer}, {Rouan}, {Samuel},
  {Santerne}, {Schneider}, {Tingley}, {Udry}, {Weingrill}, \&
  {Wuchterl}}]{CoRoT8b}
{Bord{\'e}}, P., {Bouchy}, F., {Deleuil}, M., {et~al.} 2010, \aap, 520, A66+

\bibitem[{{Bouchy} {et~al.}(2009){Bouchy}, {Moutou}, {Queloz}, \& {the CoRoT
  Exoplanet Science Team}}]{bouchy09}
{Bouchy}, F., {Moutou}, C., {Queloz}, D., \& {the CoRoT Exoplanet Science
  Team}. 2009, in IAU Symposium, Vol. 253, 129--139

\bibitem[{{Bouchy} \& {The Sophie Team}(2006)}]{bouchy06}
{Bouchy}, F. \& {The Sophie Team}. 2006, in Tenth Anniversary of 51 Peg-b:
  Status of and prospects for hot Jupiter studies, ed. {L.~Arnold, F.~Bouchy,
  \& C.~Moutou}, 319--325

\bibitem[{{Brown}(2003)}]{Brown2003}
{Brown}, T.~M. 2003, \apjl, 593, L125

\bibitem[{{Carpano} {et~al.}(2009){Carpano}, {Cabrera}, {Alonso}, {Barge},
  {Aigrain}, {Almenara}, {Bord{\'e}}, {Bouchy}, {Carone}, {Deeg}, {de La Reza},
  {Deleuil}, {Dvorak}, {Erikson}, {Fressin}, {Fridlund}, {Gondoin}, {Guillot},
  {Hatzes}, {Jorda}, {Lammer}, {L{\'e}ger}, {Llebaria}, {Magain}, {Moutou},
  {Ofir}, {Ollivier}, {Janot-Pacheco}, {P{\"a}tzold}, {Pont}, {Queloz},
  {Rauer}, {R{\'e}gulo}, {Renner}, {Rouan}, {Samuel}, {Schneider}, \&
  {Wuchterl}}]{carpano09}
{Carpano}, S., {Cabrera}, J., {Alonso}, R., {et~al.} 2009, \aap, 506, 491

\bibitem[{{Cochran} {et~al.}(1997){Cochran}, {Hatzes}, {Butler}, \&
  {Marcy}}]{16CygB}
{Cochran}, W.~D., {Hatzes}, A.~P., {Butler}, R.~P., \& {Marcy}, G.~W. 1997,
  \apj, 483, 457

\bibitem[{{Cutri} {et~al.}(2003){Cutri}, {Skrutskie}, {van Dyk}, {Beichman},
  {Carpenter}, {Chester}, {Cambresy}, {Evans}, {Fowler}, {Gizis}, {Howard},
  {Huchra}, {Jarrett}, {Kopan}, {Kirkpatrick}, {Light}, {Marsh}, {McCallon},
  {Schneider}, {Stiening}, {Sykes}, {Weinberg}, {Wheaton}, {Wheelock}, \&
  {Zacarias}}]{2mass}
{Cutri}, R.~M., {Skrutskie}, M.~F., {van Dyk}, S., {et~al.} 2003, {2MASS All
  Sky Catalog of point sources}

\bibitem[{{Deeg} {et~al.}(2009){Deeg}, {Gillon}, {Shporer}, {Rouan},
  {Stecklum}, {Aigrain}, {Alapini}, {Almenara}, {Alonso}, {Barbieri}, {Bouchy},
  {Eisl{\"o}ffel}, {Erikson}, {Fridlund}, {Eigm{\"u}ller}, {Handler}, {Hatzes},
  {Kabath}, {Lendl}, {Mazeh}, {Moutou}, {Queloz}, {Rauer}, {Rabus}, {Tingley},
  \& {Titz}}]{deeg09}
{Deeg}, H.~J., {Gillon}, M., {Shporer}, A., {et~al.} 2009, \aap, 506, 343

\bibitem[{{Dekker} \& {D'Odorico}(1992)}]{UVES}
{Dekker}, H. \& {D'Odorico}, S. 1992, The Messenger, 70, 13

\bibitem[{{Deleuil} {et~al.}(2008){Deleuil}, {Deeg}, {Alonso}, {Bouchy},
  {Rouan}, {Auvergne}, {Baglin}, {Aigrain}, {Almenara}, {Barbieri}, {Barge},
  {Bruntt}, {Bord{\'e}}, {Collier Cameron}, {Csizmadia}, {de La Reza},
  {Dvorak}, {Erikson}, {Fridlund}, {Gandolfi}, {Gillon}, {Guenther}, {Guillot},
  {Hatzes}, {H{\'e}brard}, {Jorda}, {Lammer}, {L{\'e}ger}, {Llebaria},
  {Loeillet}, {Mayor}, {Mazeh}, {Moutou}, {Ollivier}, {P{\"a}tzold}, {Pont},
  {Queloz}, {Rauer}, {Schneider}, {Shporer}, {Wuchterl}, \& {Zucker}}]{CoRoT3b}
{Deleuil}, M., {Deeg}, H.~J., {Alonso}, R., {et~al.} 2008, \aap, 491, 889

\bibitem[{{Deleuil} {et~al.}(2009){Deleuil}, {Meunier}, {Moutou}, {Surace},
  {Deeg}, {Barbieri}, {Debosscher}, {Almenara}, {Agneray}, {Granet},
  {Guterman}, \& {Hodgkin}}]{ExoDatDeleuil2009}
{Deleuil}, M., {Meunier}, J.~C., {Moutou}, C., {et~al.} 2009, \aj, 138, 649

\bibitem[{{Deleuil} {et~al.}(2011){Deleuil}, {Moutou}, \&
  {Bord{\'e}}}]{CoRoT2011EPJWC}
{Deleuil}, M., {Moutou}, C., \& {Bord{\'e}}, P. 2011, Detection and Dynamics of
  Transiting Exoplanets, St.~Michel l'Observatoire, France, Edited by
  F.~Bouchy; R.~D{\'{\i}}az; C.~Moutou; EPJ Web of Conferences, 11, 1001

\bibitem[{{Demarque} {et~al.}(2004){Demarque}, {Woo}, {Kim}, \&
  {Yi}}]{Demarque2004}
{Demarque}, P., {Woo}, J., {Kim}, Y., \& {Yi}, S.~K. 2004, \apjs, 155, 667

\bibitem[{{Derekas} {et~al.}(2011){Derekas}, {Kiss}, {Borkovits}, {Huber},
  {Lehmann}, {Southworth}, {Bedding}, {Balam}, {Hartmann}, {Hrudkova},
  {Ireland}, {Kov{\'a}cs}, {Mez{\H o}}, {Mo{\'o}r}, {Niemczura}, {Sarty},
  {Szab{\'o}}, {Szab{\'o}}, {Telting}, {Tkachenko}, {Uytterhoeven}, {Benk{\H
  o}}, {Bryson}, {Maestro}, {Simon}, {Stello}, {Schaefer}, {Aerts}, {ten
  Brummelaar}, {De Cat}, {McAlister}, {Maceroni}, {M{\'e}rand}, {Still},
  {Sturmann}, {Sturmann}, {Turner}, {Tuthill}, {Christensen-Dalsgaard},
  {Gilliland}, {Kjeldsen}, {Quintana}, {Tenenbaum}, \& {Twicken}}]{HD181068}
{Derekas}, A., {Kiss}, L.~L., {Borkovits}, T., {et~al.} 2011, Science, 332, 216

\bibitem[{{Ford}(2005)}]{Ford2005AJ}
{Ford}, E.~B. 2005, \aj, 129, 1706

\bibitem[{{Gandolfi} {et~al.}(2010){Gandolfi}, {H{\'e}brard}, {Alonso},
  {Deleuil}, {Guenther}, {Fridlund}, {Endl}, {Eigm{\"u}ller}, {Csizmadia},
  {Havel}, {Aigrain}, {Auvergne}, {Baglin}, {Barge}, {Bonomo}, {Bord{\'e}},
  {Bouchy}, {Bruntt}, {Cabrera}, {Carpano}, {Carone}, {Cochran}, {Deeg},
  {Dvorak}, {Eisl{\"o}ffel}, {Erikson}, {Ferraz-Mello}, {Gazzano}, {Gibson},
  {Gillon}, {Gondoin}, {Guillot}, {Hartmann}, {Hatzes}, {Jorda}, {Kabath},
  {L{\'e}ger}, {Llebaria}, {Lammer}, {MacQueen}, {Mayor}, {Mazeh}, {Moutou},
  {Ollivier}, {P{\"a}tzold}, {Pepe}, {Queloz}, {Rauer}, {Rouan}, {Samuel},
  {Schneider}, {Stecklum}, {Tingley}, {Udry}, \& {Wuchterl}}]{CoRoT11b}
{Gandolfi}, D., {H{\'e}brard}, G., {Alonso}, R., {et~al.} 2010, \aap, 524, A55

\bibitem[{{Guenther} {et~al.}(2009){Guenther}, {Hartmann}, {Esposito},
  {Hatzes}, {Cusano}, \& {Gandolfi}}]{30AriB}
{Guenther}, E.~W., {Hartmann}, M., {Esposito}, M., {et~al.} 2009, \aap, 507,
  1659

\bibitem[{{Hatzes} {et~al.}(2003){Hatzes}, {Cochran}, {Endl}, {McArthur},
  {Paulson}, {Walker}, {Campbell}, \& {Yang}}]{Hatzes2003}
{Hatzes}, A.~P., {Cochran}, W.~D., {Endl}, M., {et~al.} 2003, \apj, 599, 1383

\bibitem[{{Hauschildt} \& {Baron}(1999)}]{phoenix99}
{Hauschildt}, P.~H. \& {Baron}, E. 1999, Journal of Computational and Applied
  Mathematics, 109, 41

\bibitem[{{Kupka} {et~al.}(2004){Kupka}, {Landstreet}, {Sigut}, {Bildfell},
  {Ford}, {Officer}, {Silaj}, \& {Townshend}}]{HD185395}
{Kupka}, F., {Landstreet}, J.~D., {Sigut}, A., {et~al.} 2004, in IAU Symposium,
  Vol. 224, The A-Star Puzzle, ed. {J.~Zverko, J.~Ziznovsky, S.~J.~Adelman, \&
  W.~W.~Weiss}, 573--579

\bibitem[{{Mayor} \& {Mazeh}(1987)}]{mm87}
{Mayor}, M. \& {Mazeh}, T. 1987, \aap, 171, 157

\bibitem[{{Mayor} {et~al.}(2003){Mayor}, {Pepe}, {Queloz}, {Bouchy},
  {Rupprecht}, {Lo Curto}, {Avila}, {Benz}, {Bertaux}, {Bonfils}, {Dall},
  {Dekker}, {Delabre}, {Eckert}, {Fleury}, {Gilliotte}, {Gojak}, {Guzman},
  {Kohler}, {Lizon}, {Longinotti}, {Lovis}, {Megevand}, {Pasquini}, {Reyes},
  {Sivan}, {Sosnowska}, {Soto}, {Udry}, {van Kesteren}, {Weber}, \&
  {Weilenmann}}]{mayor03}
{Mayor}, M., {Pepe}, F., {Queloz}, D., {et~al.} 2003, The Messenger, 114, 20

\bibitem[{{Mazeh} {et~al.}(1997){Mazeh}, {Martin}, {Goldberg}, \&
  {Smith}}]{mazeh97}
{Mazeh}, T., {Martin}, E.~L., {Goldberg}, D., \& {Smith}, H.~A. 1997, \mnras,
  284, 341

\bibitem[{{Mazeh} {et~al.}(2006){Mazeh}, {Tamuz}, \& {North}}]{EBAS2}
{Mazeh}, T., {Tamuz}, O., \& {North}, P. 2006, \mnras, 367, 1531

\bibitem[{{Mazeh} {et~al.}(2009){Mazeh}, {Tsodikovich}, {Segal}, {Zucker},
  {Eggenberger}, {Udry}, \& {Mayor}}]{TRIMOR}
{Mazeh}, T., {Tsodikovich}, Y., {Segal}, Y., {et~al.} 2009, \mnras, 399, 906

\bibitem[{{Mazeh} {et~al.}(1995){Mazeh}, {Zucker}, {Goldberg}, {Latham},
  {Stefanik}, \& {Carney}}]{mazeh95}
{Mazeh}, T., {Zucker}, S., {Goldberg}, D., {et~al.} 1995, \apj, 449, 909

\bibitem[{{McCarthy} {et~al.}(1993){McCarthy}, {Sandiford}, {Boyd}, \&
  {Booth}}]{SANDIFORD}
{McCarthy}, J.~K., {Sandiford}, B.~A., {Boyd}, D., \& {Booth}, J. 1993, \pasp,
  105, 881

\bibitem[{{Michel} {et~al.}(2009){Michel}, {Samadi}, {Baudin}, {Barban},
  {Appourchaux}, \& {Auvergne}}]{CoRoTResponse}
{Michel}, E., {Samadi}, R., {Baudin}, F., {et~al.} 2009, \aap, 495, 979

\bibitem[{{Morton} \& {Johnson}(2011)}]{Morton2011}
{Morton}, T.~D. \& {Johnson}, J.~A. 2011, ArXiv e-prints

\bibitem[{{Pepe} {et~al.}(2000){Pepe}, {Mayor}, {Delabre}, {Kohler}, {Lacroix},
  {Queloz}, {Udry}, {Benz}, {Bertaux}, \& {Sivan}}]{pepe00}
{Pepe}, F., {Mayor}, M., {Delabre}, B., {et~al.} 2000, in Society of
  Photo-Optical Instrumentation Engineers (SPIE) Conference Series, ed. {M.~Iye
  \& A.~F.~Moorwood}, Vol. 4008, 582--592

\bibitem[{{Pepe} {et~al.}(2002){Pepe}, {Mayor}, {Galland}, {Naef}, {Queloz},
  {Santos}, {Udry}, \& {Burnet}}]{Pepe2002}
{Pepe}, F., {Mayor}, M., {Galland}, F., {et~al.} 2002, \aap, 388, 632

\bibitem[{{Perruchot} {et~al.}(2008){Perruchot}, {Kohler}, {Bouchy}, {Richaud},
  {Richaud}, {Moreaux}, {Merzougui}, {Sottile}, {Hill}, {Knispel}, {Regal},
  {Meunier}, {Ilovaisky}, {Le Coroller}, {Gillet}, {Schmitt}, {Pepe}, {Fleury},
  {Sosnowska}, {Vors}, {M{\'e}gevand}, {Blanc}, {Carol}, {Point}, {Laloge}, \&
  {Brunel}}]{SOPHIE}
{Perruchot}, S., {Kohler}, D., {Bouchy}, F., {et~al.} 2008, in Society of
  Photo-Optical Instrumentation Engineers (SPIE) Conference Series, Vol. 7014

\bibitem[{{Popper} \& {Etzel}(1981)}]{Popper1981}
{Popper}, D.~M. \& {Etzel}, P.~B. 1981, \aj, 86, 102

\bibitem[{{Raghavan} {et~al.}(2010){Raghavan}, {McAlister}, {Henry}, {Latham},
  {Marcy}, {Mason}, {Gies}, {White}, \& {ten Brummelaar}}]{Raghavan2010}
{Raghavan}, D., {McAlister}, H.~A., {Henry}, T.~J., {et~al.} 2010, \apjs, 190,
  1

\bibitem[{{Russeil}(2003)}]{Russeil2003}
{Russeil}, D. 2003, \aap, 397, 133

\bibitem[{{Sing}(2010)}]{Sing2010}
{Sing}, D.~K. 2010, \aap, 510, A21

\bibitem[{{Surace} {et~al.}(2008){Surace}, {Alonso}, {Barge}, {Cautain},
  {Chabaud}, {Deleuil}, {Fenouillet}, {Meunier}, \& {Moutou}}]{CoRoTalarm}
{Surace}, C., {Alonso}, R., {Barge}, P., {et~al.} 2008, in Society of
  Photo-Optical Instrumentation Engineers (SPIE) Conference Series, Vol. 7019,
  Society of Photo-Optical Instrumentation Engineers (SPIE) Conference Series

\bibitem[{{Tamuz} {et~al.}(2006){Tamuz}, {Mazeh}, \& {North}}]{EBAS}
{Tamuz}, O., {Mazeh}, T., \& {North}, P. 2006, \mnras, 367, 1521

\bibitem[{{Tegmark} {et~al.}(2004){Tegmark}, {Strauss}, {Blanton}, {Abazajian},
  {Dodelson}, {Sandvik}, {Wang}, {Weinberg}, {Zehavi}, {Bahcall}, {Hoyle},
  {Schlegel}, {Scoccimarro}, {Vogeley}, {Berlind}, {Budavari}, {Connolly},
  {Eisenstein}, {Finkbeiner}, {Frieman}, {Gunn}, {Hui}, {Jain}, {Johnston},
  {Kent}, {Lin}, {Nakajima}, {Nichol}, {Ostriker}, {Pope}, {Scranton},
  {Seljak}, {Sheth}, {Stebbins}, {Szalay}, {Szapudi}, {Xu}, {Annis},
  {Brinkmann}, {Burles}, {Castander}, {Csabai}, {Loveday}, {Doi}, {Fukugita},
  {Gillespie}, {Hennessy}, {Hogg}, {Ivezi{\'c}}, {Knapp}, {Lamb}, {Lee},
  {Lupton}, {McKay}, {Kunszt}, {Munn}, {O'Connell}, {Peoples}, {Pier},
  {Richmond}, {Rockosi}, {Schneider}, {Stoughton}, {Tucker}, {vanden Berk},
  {Yanny}, \& {York}}]{Tegmark}
{Tegmark}, M., {Strauss}, M.~A., {Blanton}, M.~R., {et~al.} 2004, \prd, 69,
  103501

\bibitem[{{Tokovinin}(2008)}]{Tokovinin2008}
{Tokovinin}, A. 2008, \mnras, 389, 925

\bibitem[{{Torres} {et~al.}(1995){Torres}, {Stefanik}, {Latham}, \&
  {Mazeh}}]{torres95}
{Torres}, G., {Stefanik}, R.~P., {Latham}, D.~W., \& {Mazeh}, T. 1995, \apj,
  452, 870

\bibitem[{{White} {et~al.}(2007){White}, {Gabor}, \& {Hillenbrand}}]{HD32923}
{White}, R.~J., {Gabor}, J.~M., \& {Hillenbrand}, L.~A. 2007, \aj, 133, 2524

\bibitem[{{Zucker} \& {Mazeh}(1994)}]{zm94}
{Zucker}, S. \& {Mazeh}, T. 1994, \apj, 420, 806

\bibitem[{{Zucker} {et~al.}(2003){Zucker}, {Mazeh}, {Santos}, {Udry}, \&
  {Mayor}}]{TODMOR}
{Zucker}, S., {Mazeh}, T., {Santos}, N.~C., {Udry}, S., \& {Mayor}, M. 2003,
  \aap, 404, 775

\bibitem[{{Zucker} {et~al.}(2004){Zucker}, {Mazeh}, {Santos}, {Udry}, \&
  {Mayor}}]{TODMOR2}
{Zucker}, S., {Mazeh}, T., {Santos}, N.~C., {Udry}, S., \& {Mayor}, M. 2004,
  \aap, 426, 695

\end{thebibliography}
 
\end{document}